\documentstyle[11pt,psfig]{article}

\begin{document}
\title{Identifying modes in a pulsating sdB (EC~14026) star.}
\author{M.D. Reed and the }
\author{Whole Earth Telescope Xcov 17 and 21 teams}

\begin{abstract}
We present a preliminary analysis of time series photometry on the pulsating
subdwarf B star PG1336-018  acquired
as part of two Whole Earth Telescope runs. PG1336-018 is a multi-mode
pulsator with at least 15 periods centered around 175 seconds. It is also
an HW~Vir-type
eclipsing binary with a period of 2.4 hours.

Though over 20 pulsating subdwarf B stars have been discovered, only a few have
resolved temporal spectra and still fewer have identified pulsation modes. We
analyze features unique to the eclipsing binary system of PG1336-018 as a 
means of identifying pulsation modes. 
\end{abstract}

\section{Introduction}


Pulsating sdB (known as sdBV or EC14026) stars offer us the potential to probe
the interior of Blue Horizontal Branch (or sdB for field) 
stars using asteroseismology.
PG1336-018 (hereafter PG~1336) was discovered to be a pulsating sdB star
by Kilkenny et al. (1998), with the added bonus that it is also an
eclipsing binary. The companion is a late-type $\sim$M5 dwarf,
so the stellar radii are comparable, yet the companion contributes little
to the integrated flux.  
During primary eclipse
the companion covers about half of the pulsator.

\section{Previous studies}

PG~1336 was the target of a Whole Earth Telescope (WET) campaign in April of
1999.
It was observed to pulsate in at least 15 modes (Reed \emph{et al.}, 2000) with
2 modes split by twice the orbital frequency. We wished to use the eclipse
to identify some of the remaining modes.

Pulsations manifest themselves as alternating hot
and cool regions on the star's surface. As the visible surface area changes
(during eclipse), so should the observed pulsation amplitudes. 
An analysis in which the pulsation axis was aligned with the rotation axis
 was presented in Reed, Kawaler, \& Kleinman (2001). They determined that
modes with $\ell=1$, $m=0$; $\ell=2$, $m=\pm 1$; and all $\ell=3$ modes
would \emph{only} be visible during eclipse while $\ell=2$, $m=\pm 2$ modes
would disappear during eclipse. Other modes would remain unaffected. This means
eclipses could be used to identify pulsation modes.

Unfortunately, Xcov 17 did not have the usual WET coverage and was not
optimized to study the eclipses. 
However it did prove the method to be viable and, in April, 2001, the WET 
observed PG~1336 again.

\section{Xcov 21}
During Xcov 21, $\sim$440 hours of data 
were collected from April 17-30, 2001 (a 
duty cycle of $\sim$62\%). Over 27 separate modes were identified in the 
out-of-eclipse (OoE) temporal spectrum and
this time the data were optimized to examine the
pulsations through eclipse. Though still a non-trivial task, we were able
to detect 15 peaks. As 
summarized in \S 2, we would expect to observe several modes in both the OoE
and primary eclipse data, as well as new, rotationally split modes in the
primary eclipse data. This was not the case. Of the 15 modes 
detected during primary eclipse, only 2 were observed in the
OoE data (the 2 with the largest amplitudes). 

Our first assumption was that the primary eclipse data was too
contaminated with aliases to be useful. To test this, we reduced and analyzed
data from the secondary eclipse in the same fashion as from the primary eclipse.
We readily detected 12 peaks on the temporal spectrum, 11 of which coincided
exactly (to within the errors) to those detected in the OoE data. This gave
us confidence that it was not
a reduction problem.
It more likely indicates that aligning the pulsation axis with the
rotation axis is a poor assumption.

\section{A tipped pulsation axis}
As first suggested by Margarida Cunha (private communication), 
if the tidal force dominates the Coriolis
force, the pulsation axis may align along the axis of the binary. This is 
similar to the roAp stars where the magnetic field dictates the alignment
of the pulsation axis.  To determine the effect of such an alignment, we created
simulated light curves where the pulsation axis is tipped by 90 degrees
(the inclination of the actual binary is 81 degrees) and precesses to always
point toward the companion (eclipses were ignored). Our simulated light curves and
Fourier transforms are provided in Figure\,1. Pulsation modes are indicated on
the left of each plot (as in f$\ell m$). For the light curves, phases (where a
shift of $0.5=180^o$) are dashed lines, with scales on the right of the plot
and dotted lines indicate orbital phases (PE=primary eclipse,
SE=secondary eclipse, T and A=quadrature
when the star is moving toward and away from the observer, respectively). The
dot-dashed line indicates the input frequency (corresponding to a pulsation period
of 175 seconds) of the FT, the dotted line is $1\cdot f_{orb}$ 
away from the input signal
and the dashed line is $2\cdot f_{orb}$ away from the input signal.
As indicated in the figure, \emph{every} non-radial mode undergoes at 
least two $180^o$
shifts in phase during each orbital cycle. 

%
\begin{figure}
\centerline{
\psfig{figure=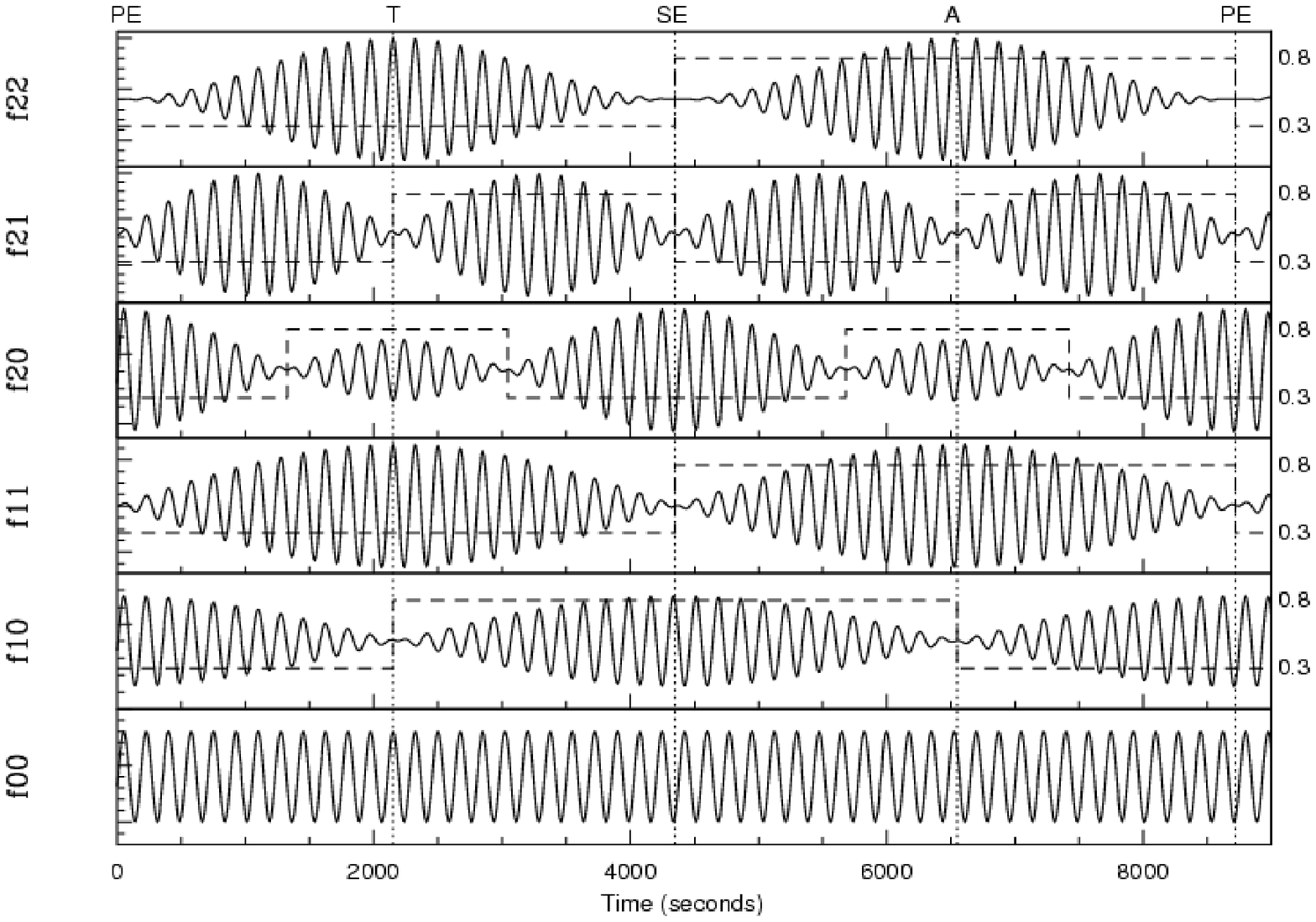,width=3in}\psfig{figure=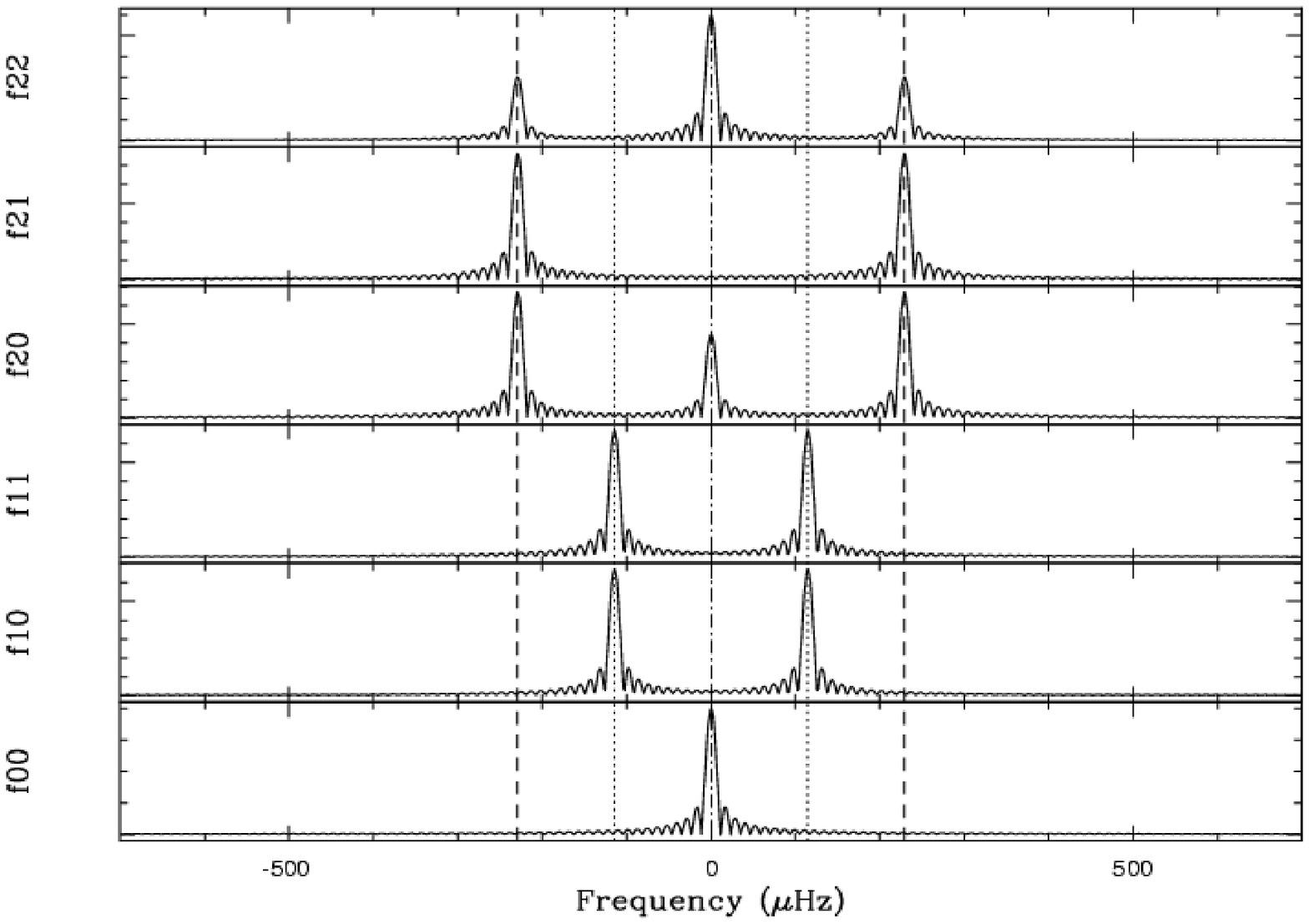,width=3in} }
\caption{Left: Simulated light curves and FTs with the pulsation axis tipped
90 degrees to the rotation axis.}
\end{figure}

Each pulsation frequency produces a complex light curve
which creates a many-peaked temporal spectrum. But this complexity may be the 
key to understanding the pulsations. PG~1336 has many peaks in the temporal
spectrum; by tipping the pulsation axis, we can create many peaks from fewer
actual pulsations. 
If the lightcurve is binned by orbital phase 
for a particular mode, then only the real peak will be present,
and predictable phase shifts for similar valued modes will be observed.

Of the 4 highest amplitude modes,
two are split by $2\cdot f_{orb}$, a signature of $\ell$=1 modes.
We divided up the light curve and binned by orbital
phase appropriate
for an $\ell$=1, $m=\pm1$ mode. We then analized each temporal spectrum
independently. In both (phase) cases, a peak midway between appeared where none
previously existed in the combined spectrum. Additionally, the phase between
the regional spectra changed, as expected for an $\ell$=1, $m=\pm1$ mode. This
provides strong indications that indeed PG~1336 has a tipped pulsation axis. The
next step will be to bin the light curve by orbital phase appropriate for
other modes and perform similar tests. Such tests will be performed in the
near future.

\subsection{Eclipses revisited}
Simulations like those in Reed, Kawaler, \& Kleinman (2001) were created to
model light curves during primary eclipse for the re-aligned pulsation axis. With
the exception of the $\ell$=2, $m=\pm2$ modes, all modes show \emph{no} effect
from the eclipse in phase and only mild changes in amplitude. Thus, if the
pulsation axis is tipped 90 degrees to the rotation axis,
amplitude during the primary eclipse is
no longer an indicator of pulsation modes.

\section{Conclusions and future work}

\begin{figure}
\centerline{
\psfig{figure=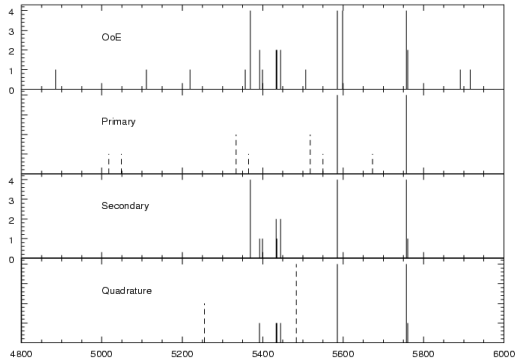,width=3in} }
\caption{Frequencies detected in the temporal spectra of PG~1336 binned in
amplitude (for clarity). Solid lines are modes present in the OoE temporal spectrum
while dashed lines indicate peaks only observed in the labeled temporal
spectra.}
\end{figure}

Figure\,3 shows the pulsations detected from various reductions of the data
for PG~1336 from the Xcov 21 WET run. More than 20 individual pulsations are
detected in the combined data set, most of which are detected in other, regional
data sets as well. The primary eclipse temporal spectrum is the most incongruent,
with only 2 modes detected in other temporal spectra. Many of the remaining 
modes are orbital separations away from modes detected in the OoE spectrum.

The temporal spectra at eclipse and quadrature provide good indications that
the pulsation axis of PG~1336 is not aligned with the rotation axis, but rather
(at least somewhat) with the tidal force from the companion. This 
likely has far broader applications than just PG~1336. Since a sizable fraction
of sdB stars are likely in binaries (Green, Liebert, \& Saffer, 2001), it is
probable (and almost a certainty for KPD~1930+2752) that other pulsators also
have pulsation axes which (somewhat) align with the tidal force of their 
companions. 

I would also like to call on theorists to help in understanding how pulsations
respond to tidal forces. It is already believed that $p$ and $g$ modes will
"feel" the tidal forces differently, but it would be helpful for observers
to have some form of parameterization so we can predict what the pulsation
axis \emph{should} be. There is also some question (at least in my mind) as to
what it means to have $m\neq 0$ modes which are not due to rotation.

\end{document}